\documentclass[twocolumn,amsmath,amssymb,showpacs]{revtex4}
\usepackage{bm}
\usepackage{graphicx}
\usepackage{epstopdf}

\begin{document}

\newcommand{\uu}[1]{\underline{#1}}
\newcommand{\pp}[1]{\phantom{#1}}
\newcommand{\be}{\begin{eqnarray}}
\newcommand{\ee}{\end{eqnarray}}
\newcommand{\ve}{\varepsilon}
\newcommand{\vs}{\varsigma}
\newcommand{\Tr}{{\,\rm Tr\,}}
\newcommand{\pol}{\frac{1}{2}}

\title{
Relativity of arithmetic as a fundamental symmetry of physics}
\author{Marek Czachor}
\affiliation{
Katedra Fizyki Teoretycznej i Informatyki Kwantowej,
Politechnika Gda\'nska, 80-233 Gda\'nsk, Poland\\
Centrum Leo Apostel (CLEA),
Vrije Universiteit Brussel, 1050 Brussels, Belgium
}

\begin{abstract}
Arithmetic operations can be defined in various ways, even if one assumes commutativity and associativity of addition and multiplication, and distributivity of multiplication with respect to addition. In consequence, whenever one encounters `plus' or `times' one has certain freedom of interpreting this operation. This leads to some freedom in definitions of derivatives, integrals and, thus, practically all equations occurring in natural sciences. A change of realization of arithmetic, without altering the remaining structures of a given equation, plays the same role as a symmetry transformation. An appropriate construction of arithmetic turns out to be particularly important for dynamical systems in fractal space-times. Simple examples from classical and quantum, relativistic and nonrelativistic physics are discussed, including the eigenvalue problem for a quantum harmonic oscillator. It is explained why the change of arithmetic is not equivalent to the usual change of variables, and why it may have implications for the Bell theorem.
\end{abstract}
\maketitle

\section{Introduction}

Symmetries of physical systems can be rather obvious or very abstract. Lorentz transformations, discovered as a formal symmetry of Maxwell's equations, seemed abstract until their physical meaning was understood by Einstein. Theory of group representations, the cornerstone of quantum mechanics and field theory, had its roots in Lie's studies of abstract symmetries of differential equations. It  has taught us that differences in mathematical realizations of a symmetry may directly reflect physical differences.

Einstein's relativity, gauge invariance, Noether's theorems, Darboux-B\"acklund transformations, or supersymmetry are prominent examples of symmetry principles in physics.
In the paper we discuss a new type of principle, occurring in any physical theory: The symmetry of mathematical equations under modifications of arithmetic operations, the induced modifications of derivatives and integrals included. Similarly to other physical symmetries, the symmetry maintains the form of relevant equations, but may possess different mathematical realizations.

The approach developed in this paper is not in any sense close to those generalizations of physics that involve $p$-adic mathematics \cite{p-ad1,p-ad2}, although the context of fractals and the Cantor set might create such an impression. Another formalism that can be easily confused with what we do is based on `arithmetic dynamics' \cite{ArDyn} which involves the so-called linear conjugates $F_f=f^{-1}\circ F\circ f$ (which indeed will play a role here) but where the calculus is defined in the standard way.

Our formalism is a by-product of attempts of understanding the dynamics of fractals, but turned out to have more general consequences. In particular, it opens some new perspectives on quantum mechanics of processes occurring in sets of Lebesgue measure zero, invisible from the point of view of standard quantum mechanics, leading to phenomena analogous to dark energy, `coming out of nowhere'.

In a wider context, what we do is closest to the works of Burgin \cite{Burgin} on non-Diophantine arithmetic, and Benioff on physical consequences of  rescaled multiplication \cite{Benioff,Benioff1}. However, neither Burgin nor Benioff made the essential step of formulating a non-Diophantine calculus.

\section{From arithmetic to calculus}

To begin with, let us consider a bijection $f:X\to Y\subset\mathbb{R}$, where $X$ is some set. The map $f$ allows us to define addition, multiplication, subtraction, and division in $X$,
\be
x\oplus y &=& f^{-1}\big(f(x)+f(y)\big),\nonumber\\
x\ominus y &=& f^{-1}\big(f(x)-f(y)\big),\nonumber\\
x\odot y &=& f^{-1}\big(f(x)f(y)\big),\nonumber\\
x\oslash y &=& f^{-1}\big(f(x)/f(y)\big).\nonumber
\ee
One easily verifies the standard properties \cite{Proofs}: (1) associativity $(x\oplus y)\oplus z=x\oplus (y\oplus z)$,
$(x\odot y)\odot z=x\odot (y\odot z)$, (2) commutativity $x\oplus y=y\oplus x$, $x\odot y=y\odot x$, (3) distributivity
$(x\oplus y)\odot z=(x\odot z)\oplus (y\odot z)$. In fact, this is precisely an example of a non-Diophantine arithmetic in the sense of \cite{Burgin}. The rescaled-multiplication formalism of Benioff \cite{Benioff} corresponds to $f(x)=px$, $p\neq 0$, and in its generalized versions \cite{Benioff1} to bundles of functions $f_X(x)$ where $X$ is a space-time point.

Elements $0,1\in X$ are defined by $0\oplus x=x$, $1\odot x=x$, which implies $f(0)=0$, $f(1)=1$.
One further finds $x\ominus x=0$, $x\oslash x=1$, as expected \cite{rem1}. In general, it is better to define subtraction independently of addition since it may happen that $f(-x)$ is undefined. An important nontrivial example of $f$ is provided by the Cantor function \cite{CantorF}, or more precisely the Cantor-line function defined below, where $X$ is the Cantor subset of $[0,1]$ and $Y=[0,1]$. If $0\ominus x$ exists, one can denote it by $\ominus x$.

Practically the only difference between $\oplus$, $\ominus$, $\odot$, $\oslash$ and $+$, $-$, $\cdot$, and $/$ is that in general multiplication is not just a repeated addition: Typically $x\oplus x\neq 2\odot x$. Multiplication and addition are now truly independent.

Having all these arithmetic operations one can define a derivative of a function $A:X\to X$,
\be
\frac{d_f A(x)}{d_fx}
&=&
\lim_{h\to 0}\Big(A(x\oplus h)\ominus A(x)\Big)\oslash h,\label{der}
\ee
satisfying
\be
\frac{d_f A(x)\odot B(x)}{d_fx}
&=&
\frac{d_f A(x)}{d_fx}\odot B(x)
\oplus
A(x)\odot \frac{d_fB(x)}{d_fx},\nonumber\\
\frac{d_f A(x)\oplus B(x)}{d_fx}
&=&
\frac{d_f A(x)}{d_fx}
\oplus
\frac{d_f B(x)}{d_fx},
\nonumber\\
\frac{d_f A[B(x)]}{d_fx}
&=&
\frac{d_f A[B(x)]}{d_f B(x)}
\odot
\frac{d_f B(x)}{d_fx}.\nonumber
\ee

Now consider functions $F:Y\to Y$ and $F_f:X\to X$ related by
\be
F_f(x) = f^{-1}\Big(F\big(f(x)\big)\Big).\label{F_f}
\ee
Employing (\ref{der}) and the fact that $f(0)=0$ one finds
\be
\frac{d_f F_f(x)}{d_fx}
&=&
f^{-1}\Big(F'\big(f(x)\big)\Big),\label{derf}
\ee
where $F'(y)=dF/dy$ is the usual derivative in $Y$, defined in terms of $+$, $-$, $\cdot$, and $/$.
It is extremely important to note that (\ref{derf}) has been derived with no need of differentiability of $f$. $f(0)=0$ is enough to obtain a well defined derivative.
(\ref{derf}) is not the standard formula known for composite functions since no derivatives of $f$ occur \cite{d/dx}.

An integral is defined so that the fundamental laws of calculus,
\be
\int_a^b \frac{d_f A(x)}{d_fx}\odot d_fx = A(b)\ominus A(a),\nonumber
\ee
and
\be
\frac{d_f}{d_f x}\int_a^x A(x')\odot d_fx' = A(x),
\ee
hold true. The explicit form reads
\be
\int_a^b F_f(x)\odot d_fx = f^{-1}\left(\int_{f(a)}^{f(b)}F(y)dy\right),\label{int}
\ee
where $\int F(y)dy$ is the standard (say, Lebesgue) integral in $\mathbb{R}$.

\section{Correspondence principle}

Let us consider the two plots of the same function $f(x)$ shown in Fig.~1. The right one is practically indistinguishable from $f(x)=x$, which means that numbers that are sufficiently large will satisfy an effectively Diophantine arithmetic and calculus, with $x\oplus y\approx x+y$, $x\odot y\approx xy$, $d_f/d_fx\approx d/dx$, etc. Functions $f$ that lead to such a {\it correspondence principle\/} can be characterized by
\be
\lim_{\lambda\to\infty}\frac{f(\lambda x)}{\lambda}=x.
\ee
\begin{figure}
\includegraphics[width=4cm]{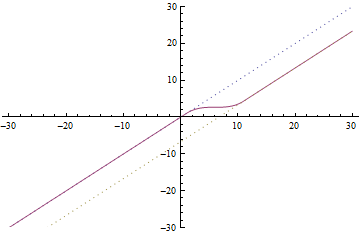}
\includegraphics[width=4cm]{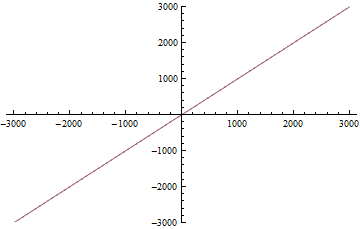}
\caption{Diophantine correspondence principle. Left: $f(x)$ that interpolates between $f(x)=x$ (for $x\leq 1$) and $f(x)=x-20/3$ (for $x\geq 10$). In the transition region $f(x)=(-16 + 108 x - 18 x^2 + x^3)/75$. $f:\mathbb{R}\to\mathbb{R}$ is bijective and satisfies $f(0)=0$, $f(1)=1$. Arithmetic defined in terms of $f$ is non-Diophantine, but is well approximated by the standard Diophantine case $f(x)=x$ if $x$ is large enough. Right: The same $f(x)$ but plotted in a different scale. Here $f(x)$ is practically indistinguishable from $f(x)=x$ and thus the resulting arithmetic is effectively Diophantine.}
\end{figure}

Anticipating results of the next section let us show, in Fig.~2, the exponential function resulting from the solution of (\ref{dA/dt=A}) and with arithmetic defined by $f(x)$ from Fig.~1. As one can see, the resulting exponent involves two standard exponential regimes separated by a `non-exponential' part. Still, one has to keep in mind that this awkward looking function does nevertheless satisfy the fundamental property (\ref{exp(x+y)}), so this is just {\it the\/} exponential function corresponding to the non-Diophantine arithmetic and calculus implied by this concrete $f$. The exponent $\exp_f$ from Fig.~2 has striking similarity to the expansion factor occurring in the $\Lambda$CDM standard model of cosmology. The two `standard' exponential regimes thus can be termed the inflationary and dark-energy parts of $\exp_f x$.

\begin{figure}
\includegraphics[width=6cm]{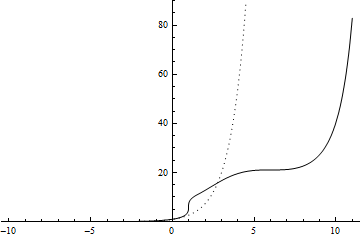}
\caption{Ordinary $e^x$ (dotted) versus $\exp_f x$ defined in terms of $f$ from Fig.~1. Note the `inflation' and `dark energy' parts of $\exp_f x$.}
\end{figure}

\section{Examples}

\subsection{Exponential function}

To understand why functions of the form (\ref{F_f}) are so essential let us solve the differential equation
\be
\frac{d_f A(x)}{d_fx}
&=&
A(x), \quad A(0)=1 \label{dA/dt=A}
\ee
by assuming that
$
A(x)
=
\oplus_{n=0}^\infty a_n\odot x^{\odot n}
$,
where $x^{\odot n}=x\odot\dots\odot x$ ($n$ times). Then, comparing term by term, one finds the unique solution
\be
A(x)=f^{-1}\big(e^{f(x)}\big)=\exp_f x,\label{exp_f}
\ee
fulfilling
\be
\exp_f (x\oplus y)=\exp_f x \odot\exp_f y.\label{exp(x+y)}
\ee
Its inverse is
$\ln_f x = f^{-1}\big(\ln f(x)\big)$,
$\ln_f (x\odot y) = \ln_f x\oplus \ln_f y$.

An instructive exercise is to take $f(x)=x^3$, $f^{-1}(x)=\sqrt[3]{x}$. Then $x\odot y=xy$, $x\oplus y=\sqrt[3]{x^3+y^3}$, and $\exp_f=e^{x^3/3}$ solves (\ref{dA/dt=A}). Transformation $f(x)=x\to f(x)=x^3$, in spite of its nonlinearity, keeps (\ref{dA/dt=A}) unchanged. Accordingly, this is a symmetry of (\ref{dA/dt=A}), a fact explaining the title of the paper.

\subsection{Classical harmonic oscillator: Trigonometric functions}

As our next example consider a classical harmonic oscillator
\be
\frac{d^2_f x(t)}{d_ft^2}
&=&
\frac{d_f}{d_ft}
\frac{d_f x(t)}{d_ft}
=
\ominus \omega^{\odot 2}\odot x(t)\nonumber
\ee
where $\omega^{\odot 2}=\omega\odot\omega$. Setting
$
x(t)
=
\oplus_{n=0}^\infty a_n\odot t^{\odot n},
$
one obtains
\be
x(t)&=& C_1\odot \sin_f(\omega\odot t)\oplus C_2\odot \cos_f(\omega\odot t)\nonumber
\ee
where
\be
\sin_f x = f^{-1}\big(\sin f(x)\big),\quad
\cos_f x = f^{-1}\big(\cos f(x)\big),\nonumber
\ee
and $C_1$, $C_2$ are constants.

An instructive exercise is to plot phase-space trajectories of the harmonic oscillator corresponding to various choices of $f$. Fig.~3 shows the trajectories for the Cantor-line function, defined below, and $f(x)=x^n$, with $n=1,3,5$. All these trajectories represent a classical harmonic oscillator that satisfies the usual law of `force oppositely proportional to displacement', with conserved energy `$\dot x^2+\omega^2x^2$',  but with different meanings of `plus' and `times'. The resulting trigonometric functions are essentially the chirp signals \cite{chirp} known from signal analysis.

\subsection{Fractals: Cantor set}

If one still has impression that what we do is just standard physics in nonstandard coordinates, consider the  problem of a fractal Universe of dimension $4-\epsilon$, analogous to the one arising in causal dynamical triangulation theory \cite{Ambjorn}. Our physical equations have to be formulated in terms of notions that are intrinsic to the Universe, but what should be meant by a velocity, say? We have to subtract positions and divide by time, but we have to do it in a way that is intrinsic to the Universe we live in. Moreover, from our perspective positions and flow of time seem continuous even if they would appear  discontinuous from an exactly 4-dimensional perspective. We should not make the usual step and turn to fractional derivatives \cite{Podlubny}, since for inhabitants of $(4-\epsilon)$-dimensional Universe the velocity is just the first derivative of position with respect to time, and not some derivative of order $0<\alpha<1$.

As usual, Cantor-like sets and Cantor-type functions provide a rich source of highly nontrivial examples \cite{Thirring,Amrein}. A simple model of $(4-\epsilon)$-dimensional space-time is the Cartesian product of four Cantor dusts, appropriately extended to the whole of $\mathbb{R}$. Cantor dusts are easy to work with and possess certain mathematical universality \cite{Edgar}, but can be defined in different ways. Here we need a precisely constructed fractal  $X$ and a bijection $f:X\to Y$. Let us start with the right-open interval $[0,1)\subset \mathbb{R}$, and let the (countable) set $Y_2\subset[0,1)$ consist of those numbers that have two different binary representations. Denote by $0.t_1t_2\dots$ a ternary representation of some $x\in [0,1)$. If $y\in Y_1=[0,1)\setminus Y_2$ then $y$ has a unique binary representation, say $y=0.b_1b_2\dots$. One then sets $f^{-1}(y)=0.t_1t_2\dots$, $t_j=2b_j$. Let $y=0.b_1b_2\dots=0.b'_1b'_2\dots$ be the two representations of $y\in Y_2$. We define $f^{-1}(y)=\min\{0.t_1t_2\dots,0.t'_1t'_2\dots\}$, where $t_j=2b_j$, $t'_j=2b'_j$. We have therefore constructed an injective map $f^{-1}:[0,1)\to [0,1)$. The triadic Cantor-like set is defined as the image $C_{[0,1)}=f^{-1}\big([0,1)\big)$, and $f:C_{[0,1)}\to[0,1)$, $f=(f^{-1})^{-1}$, is the required bijection between $C_{[0,1)}$ and the interval. For example, $1/2\in Y_2$ since $1/2=0.1_2=0.0(1)_2$. We find
$
f^{-1}(1/2)=\min\{0.2_3=2/3,0.0(2)_3=1/3\}=1/3.
$
Accordingly, $1/3\in C_{[0,1)}$ while $2/3\notin C_{[0,1)}$. $C_{[0,1)}$ is not exactly the standard Cantor set, but all irrational elements of the Cantor set belong to $C_{[0,1)}$ (an irrational number has a unique binary form), together with some rational numbers such as $1/3$. Note further that $0\in C_{[0,1)}$, with $f(0)=0$. We could proceed analogously with $1\notin [0,1)$, since $1=1.(0)_2=0.(1)_2$ possesses two binary representations with
$\min\{2.(0)_3,0.(2)_3\}=\min\{2,1\}=1$. However, instead of including 1 in $C_{[0,1)}$, let us shift $C_{[0,1)}$ to the right by 1, thus obtaining
$C_{[1,2)}$. Proceeding in this way we construct a fractal $C=\cup_{k\in\mathbb{Z}}C_{[k,k+1)}$, and the bijection $f:C\to \mathbb{R}$. Explicitly, if $x\in C_{[0,1)}$, then $x+k\in C_{[k,k+1)}$, and $f(x+k)=f(x)+k$ by definition. Let us call $C$ the Cantor line, and $f$ the Cantor-line function.

\begin{figure}
\includegraphics[width=7cm]{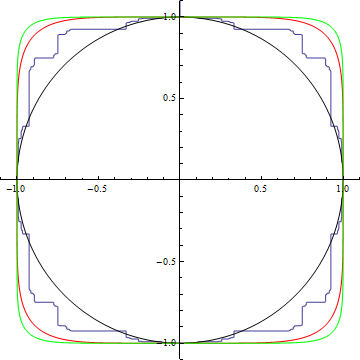}
\caption{Phase-space trajectories of the harmonic oscillator with $\omega=1$ and $f(x)=x$ (black), $f(x)=x^3$ (red), $f(x)=x^5$ (green), and the Cantor-line function (blue). Taking $f(x)=x^n$ with sufficiently large $n$ we would find a dynamics looking like a motion along a square. Note that the oscillator is described by the standard liner equation, even if we take a highly nonlinear function such as $f(x)=x^n$. This shows that a change of $f$ is more than a change of variables.}
\end{figure}

The integral we have defined is not equivalent to the fractal measure. Indeed, the fractal measure of the Cantor set embedded in an interval of length $L$ is $L^D$, where $D=\log_32$. Thus, for $L=1/3$ one finds $L^D=1/2$. Since segments $[0,1/3]$ and $[2/3,1]$ both have $L=1/3$ they both have the same $D$-dimensional volume equal $1/2$. Taking $F_f(x)=1$ we find
\be
\int_a^b d_fx=\int_a^b \frac{d_f x}{d_fx}\odot d_fx = f^{-1}\big(f(b)-f(a)\big),\nonumber
\ee
and
$\int_0^{1/3} d_fx=1/3$, $\int_{1/3}^{1} d_fx=1/3$, $\int_{0}^{1} d_fx=(1/3)\oplus(1/3)=1$.
\begin{figure}
\includegraphics[width=7cm]{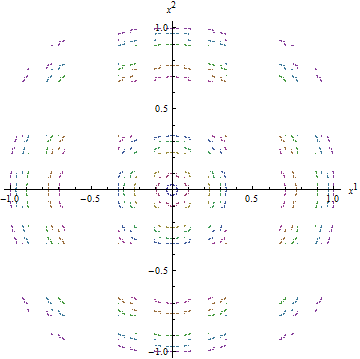}\\
\caption{A new type of rotational symmetry. Circles in Cantorian plane $C^2$ for ten different radii. The rotational symmetry is hidden in the sense that in order to reveal it one has to control the arithmetic intrinsic to the fractal set $C$.}
\end{figure}

\subsection{Rotations in fractal plane}

Now let us switch to higher dimensional examples. First consider the plane, i.e. the Cartesian product of two lines. One checks that
$
\sin_f^{\odot 2} x\oplus \cos_f^{\odot 2} x
=
1$,
$\cosh_f^{\odot 2} x \ominus \sinh_f^{\odot 2} x
=
1$. It is an appropriate place to stress that our approach does not seem to be related to exponential, trigonometric, and hyperbolic functions defined (in various ways) in the context of time-scales dynamics \cite{Hilger,Bohner,Cieslinski}, or in non-extensive thermodynamics \cite{Tsallis,Naudts}, but there are links to Kolmogorov-Nagumo averages employed in R\'enyi information theory \cite{Naudts,Renyi} (see below).
Functions $\sin_f$, $\cos_f$, $\sinh_f$, $\cosh_f$,  satisfy the basic standard formulas such as
$$
\sin_f(a\oplus b)
=
\sin_f a\odot\cos_f b
\oplus
\cos_f a\odot \sin_f b
$$
and the like, so
\be
x' &=& x \odot \cos_f \alpha\oplus y \odot \sin_f \alpha,\nonumber\\
y' &=& y \odot \cos_f \alpha \ominus x \odot \sin_f \alpha,\nonumber
\ee
defines a rotation. The rotation satisfies the usual group composition rule, a fact immediately implying that one can work with generalized-arithmetic matrix equations.

\subsection{Fractal Minkowski space}

In an analogous way one arrives at Lorentz transformations
in Cantorian Minkowski space, the Cartesian product of four Cantor lines with the invariant form
$$
x_\mu\odot x^\mu=x_0^{\odot 2}\ominus x_1^{\odot 2}\ominus x_2^{\odot 2}\ominus x_3^{\odot 2}.
$$
Such Lorentz transformations are unrelated to those occurring in exactly 4-dimensional fractal space-time of scale relativity \cite{Nottale}. In $1+1$ dimensions, employing,
\be
x'^0 &=& x^0 \odot \cosh_f \alpha\ominus x^1 \odot \sinh_f \alpha,\nonumber\\
x'^1 &=& \ominus x^0 \odot \sinh_f \alpha \oplus x^1 \odot \cosh_f \alpha ,\nonumber
\ee
and $x'^1=0$, one finds
$$
\beta=x^1\oslash x^0=\tanh_f\alpha.
$$
In consequence, the fact that $f(1)=1$ sets the limit $|\beta|\leq 1$ for maximal velocity independently of the choice of $f$. In principle, problems such as clock synchronization, composition of velocities, or the twin paradox, may lead to direct experimental tests of $f$ and some insights into a putative fractal structure of space-time.

\subsection{Harmonic oscillator and quantum mechanics}

Arithmetic of complex numbers requires some care. One should not just take $f: \mathbb{C}\to \mathbb{C}$ due to the typical multi-valuedness of $f^{-1}$ and the resulting ill-definiteness of $\oplus$ and $\odot$. Definition of $i$ as a $\pi/2$ rotation also does not properly work since one cannot guarantee a correct behaviour of $i^{\odot n}$ for a general $f$. The correct solution is the simplest one: One should treat complex numbers as pairs of reals satisfying the following arithmetic
\be
(x,y)\oplus (x',y') &=& (x\oplus x',y\oplus y'),
\nonumber\\
(x,y)\odot (x',y') &=& (x\odot x'\ominus y\odot y',y\odot x' \oplus x\odot y'),
\nonumber\\
i &=& (0,1),\nonumber
\ee
supplemented by conjugation $(x,y)^* = (x,\ominus y)$. 
The modulus reads
\be
|z|^{\odot 2}
&=&
(x,y)\odot (x,\ominus y)= (x^{\odot 2}\oplus y^{\odot 2},0).
\ee
As stressed in \cite{Rudin}, the resulting complex structure is just the standard one, but no mysterious `imaginary number' is employed.

In this way we have arrived at quantum mechanics. The most obvious example is the eigenvalue problem for a 1-dimensional harmonic oscillator.
Consider
\be
\hat H_f \psi_f(x) &=& \ominus\alpha^{\odot 2}\odot\frac{d^2_f\psi_f(x)}{d^2_f x}\oplus \beta^{\odot 2}\odot x^{\odot 2}\odot\psi_f(x)
\nonumber\\
&=& E_f\odot\psi_f(x),\label{Hpsi}
\ee
where $\alpha$, $\beta$ are parameters.
Since $\psi_f(x)=f^{-1}\Big(\psi\big(f(x)\big)\Big)$, then acting with $f$ on both sides of (\ref{Hpsi}) one finds
\be
f(E_f)\psi(y)
&=&
-f(\alpha)^{2}\psi''(y)+f(\beta)^{2}y^{2}\psi(y)
\ee
where $y=f(x)$. This is the standard harmonic oscillator equation
with eigenvalues $f(E_{n,f})=f(\alpha)f(\beta)(2n+1)$, and thus
\be
E_{n,f} &=& f^{-1}\big(f(\alpha)f(\beta)(2n+1)\big)\label{E_{n,f}}.
\ee
The eigenvalues of the quantum harmonic oscillator {\it explicitly depend on the choice of arithmetic\/}.
The normalized ground state is
\be
\psi_{0f}(x) &=& f^{-1}\Bigg(\left(\frac{f(\beta)}{\pi f(\alpha)}\right)^{1/4} e^{-\frac{f(\beta)f(x)^2}{2f(\alpha)}}\Bigg),\nonumber
\ee
with $E_{0,f}=\alpha\odot\beta$. The normalization means that 
\be
\int_{-\infty}^\infty d_fx\odot |\psi_{0f}(x)|^{\odot 2}=1.
\ee
The excited states can be derived in the usual way.

There are two peculiarities of the resulting quantum mechanics one should be aware of. First of all, if $f$ is a Cantor-like function representing a fractal whose dimension is less than 1, then the real-line Lebesgue measure of the fractal is zero. Keeping in mind that states in quantum mechanics are represented by equivalence classes of wave functions that are identical up to sets of measure zero, we can remove the Cantor line from $\mathbb{R}$ without altering standard quantum mechanics. Having removed the Cantor line $C$ from $\mathbb{R}$ we still can do ordinary quantum mechanics on $\mathbb{R}\setminus C$, whereas $C$ itself can become a universe for its own, Cantorian theory. Removing $C$ from $\mathbb{R}$ does not mean that we impose some fractal-like boundary conditions or consider a Schr\"odinger equation with a delta-peaked potential of Cantor-set support \cite{FractalQM}. We just use the freedom to modify wave functions on sets of measure zero. So we can keep the standard Gaussian $f(x)=x$ ground state on $\mathbb{R}\setminus C$, and employ the Cantorian $\psi_{0f}(x)$ on $C$. According to quantum mechanics the resulting wave function belongs to the same equivalence class as the usual Gaussian, and thus represents the same state. However, now the energy is $\hbar\omega/2+\alpha\odot \beta$, with $\alpha\odot \beta$ `appearing from nowhere'.  The analogy to dark energy is evident. The additional energy is a real number so it can be added to $\hbar\omega/2$, similarly to many other energies that occur in physics and are additive in spite of unrelated origins.

The second subtlety concerns physical dimensions of various quantities occurring in $f$-generalized arithmetic. Even the simple case of $\omega\odot t$ may imply a necessity of dimensionless $\omega$ and $t$ if $f$ is sufficiently nontrivial. Functions of the form $f(x)=x^q$, $q\in \mathbb{R}_+$, are in this respect exceptional since then $x\odot y=xy$, and one can work with dimensional quantities. For example, the harmonic oscillator for $f(x)=x^q$ has energy levels
\be
E_{n,f}=\frac{\hbar\omega}{2}(2n+1)^{\frac{1}{q}}.
\ee
In the formalism of Benioff \cite{Benioff,Benioff1}, which employs a linear $f(x)=px$, one obtains
\be
E_{n,f}=p\hbar\omega(n+1/2).
\ee
In general, for example in a fractal space-time, we have to work with dimensionless variables $x$ in order to make $f(x)$ meaningful. It is thus simplest to begin with reformulating all the `standard' theories in dimensionless forms, similarly to $c=1$ and $\hbar=1$ conventions often employed in relativity and quantum theory.

\subsection{Probability and entropy}

Quantum mechanics has brought us to the issue of probability. An appropriate normalization is $\oplus_k p_k=1$ which, in virtue of $f(1)=1$, implies $\sum_kf(p_k)=\sum_k P_k=1$. We automatically obtain two coexisting but inequivalent sets of probabilities, in close analogy to probabilities $P_k$ and escort probabilities $p_k=P_k^q$ occurring in generalized statistics and multifractal theory \cite{Tsallis, Naudts}.
Averages
\be
\langle a\rangle_f
=
\oplus_k p_k\odot a_k
=
f^{-1}\Big(\sum_k P_kf(a_k)\Big),\nonumber
\ee
have the form of Kolmogorov-Nagumo averages \cite{Naudts,CN,Jizba}, which implies the usual bounds $a_{\rm min}\leq \langle a\rangle_f\leq a_{\rm max}$.
As a by-product of the construction we have arrived at an entirely new interpretation of the Kolmogorov-Nagumo averaging: These are just the ordinary averages, but evaluated in a non-Diophantine arithmetic.

From the point of view of modified arithmetic the constraints one should impose on escort probabilities and Kolmogorov-Nagumo averages are, though, completely different from those employed in nonextensive statistics and R\'enyi's information theory \cite{Renyi}, provided instead of additivity one has $\oplus$-additivity in mind. R\'enyi's
\be
f(x)=\frac{2^{(1-q)x}-1}{2^{1-q}-1}\nonumber
\ee
can be replaced by a much wider class of $f$s.
In consequence, instead of R\'enyi's entropy one finds a general class of entropies parametrized by $f$,
\be
S_f
=
\oplus_k p_k\odot \ln_f (1\oslash p_k)
=
f^{-1}\Big(\sum_k P_k \ln P_k\Big),\nonumber
\ee
where $P_k=f(p_k)$.

The standard proof leads to the analogue
\be
\big|\langle AB\rangle_f\oplus \langle AB'\rangle_f\oplus \langle A'B\rangle_f\ominus \langle A'B'\rangle_f\big|\leq f^{-1}(2),\label{CCHSH}
\ee
of the CHSH inequality \cite{CHSH}.
Note that, in general, $f^{-1}(2)\neq 2$. Incidentally, for our choice of the Cantor-line function, with $f(x+k)=f(x)+k$, $k\in \mathbb{Z}$, one
finds the standard CHSH bound $f^{-1}(2)=2$.

It cannot be {\it a priori\/} excluded that local hidden variables satisfy a non-Diophantine variant of arithmetic. (\ref{CCHSH}) indicates that Bell's theorem contains a loophole of purely arithmetic origin.

\section{Remarks}

The modified calculus is as simple as the one one knows from undergraduate education. What may be nontrivial is to find $f$ if $X$ is a sufficiently `strange' object. The case of the Cantor line was relatively obvious, but the choice of $f$ may be much less evident if $X$ is a multifractal or a higher-dimensional fractal. Such more complicated examples are explicitly discussed in the sequel to the present paper \cite{ACK}.

In order to conclude, let us return to Fig.~3. All the phase-space trajectories represent the {\it same\/} physical system: A harmonic oscillator satisfying the Newton equation $d^2x/dt^2=-\omega^2x$, with the same physical parameters for each of the trajectories. So how come the trajectories are different? The answer is: Because the very form of Newton's equation does not tell us what should be meant by `plus' or `times'. This observation extends to any theory that employs arithmetic of real numbers. It would not be very surprising if some alternative arithmetic proved essential for Planck-scale physics, where fractal space-time is expected \cite{Benedetti,Modesto,Nicolini,Calcagni}, or to biological modeling where fractal structures are ubiquitous. Links of $f$-generalized trigonometric functions to chirps suggest that $f$s even much more `ordinary' than fractal functions may also lead to nontrivial applications. Spectrum of the harmonic oscillator is explicitly $f$-dependent, a fact showing that different versions of arithmetic lead to non-equivalent descriptions of physics.

Equation (\ref{dA/dt=A}) and its solution (\ref{exp_f}) illustrate another important difference between the change of arithmetic and the usual change of variables, in spite of the misleading similarity of some formulas. Namely, note that a change of $f$ from, say, $f(x)=x$ to an arbitrary $f$ does not influence the {\it linearity\/} of (\ref{dA/dt=A}), even for a nonlinear $f$. This should be contrasted with nonlinear gauge transformations associated with Doebner-Goldin equations \cite{DG1,DG2,DG3,DGMC}, that map a linear Schr\"odinger equation into a nonlinear one. Such a preservation of form is typical of all differential equations formulated within the generalized non-Diophantine paradigm. Another reason why one should not confuse the transformation $F\to f^{-1}\circ F\circ f=F_f$ with a gauge transformation is the resulting rule for  $dF/dx\to d_fF_f/d_fx$ which is completely different from what one finds in gauge theories. Modification of arithmetic is a symmetry operation that works at a much more primitive level than a gauge transformation.

\section*{Acknowledgments}

I am indebted to D. Aerts, P. Benioff, M. Burgin, J. L. Cie\'sli\'nski, M. Kuna and J. Naudts for discussions and critical comments.


\begin{thebibliography}{99}

\bibitem{p-ad1}V. S. Vladimirov, I. V. Volovich, and E. I. Zelenov, {\it P-adic analysis and mathematical physics\/}, World Scientific, Singapore (1994).
\bibitem{p-ad2}S. Albeverio, A. Yu. Khrennikov, and V. M. Shelkovich, {\it Theory of p-adic Distributions:
Linear and Nonlinear Models\/}, Cambridge University Press, Cambridge (2010).
\bibitem{ArDyn}J. H. Silverman, {\it The Arithmetic of Dynamical Systems\/}, Springer, New York (2007).
\bibitem{Burgin}M. Burgin, {\it Non-Diophantine Arithmetics\/}, Ukrainian Academy of Information Sciences, Kiev (1997) (in Russian).
Introduction to projective arithmetics, arXiv:1010.3287 [math.GM] (2010).
\bibitem{Benioff}P. Benioff, Int. J. Theor. Phys. {\bf 50}, 1887 (2011).
\bibitem{Benioff1}P. Benioff, Principal fiber bundle description of number scaling for scalars and vectors: Application to gauge theory, arXiv:1503.05600 [math-ph] (2015).
\bibitem{Proofs}Detailed proofs can be found in the preprint version of the paper, see arXiv:1412.8583v2 [math-ph].
\bibitem{rem1}Keeping the same symbols for $0,1\in X$ and $0,1\in \mathbb{R}$ will not lead to ambiguities. However, in the approach of Benioff \cite{Benioff,Benioff1}, where $f(x)=px$, $p\neq 0$, one finds $f^{-1}(1)=1/p$ and thus $1/p=1_p$ is the unit element of multiplication. So here one has to carefully distinguish between $1_p\in X$ and $1\in Y$.
\bibitem{CantorF}O. Dovgosheya, O. Martiob, V. Ryazanova, and M. Vuorinenc, Expo. Math. {\bf 24}, 1 (2006).
\bibitem{d/dx}To appreciate the difference between $d/dx$ and $d_f/d_fx$ take $f(x)=x^3$ and let $\sin_f x=f^{-1}\big(\sin f(x)\big)=\sqrt[3]{\sin (x^3)}$. Then
$$
\frac{d_f\sin_f x}{d_fx}=\sqrt[3]{\cos (x^3)},
\quad
\frac{d \sin_f x}{dx}
=
\frac{x^2 \cos (x^3)}{\sin ^{\frac{2}{3}}(x^3)}.
$$
The formalism from \cite{ArDyn} employs functions of the form $F_f(x)=f^{-1}\big(F f(x)\big)$ but the derivatives are given by $d/dx$ and not by $d_f/d_fx$. A consistent application of generalized arithmetic leads to a simpler formalism.
\bibitem{chirp}R. L. Easton Jr., {\it Fourier Methods in Imaging\/}, Wiley, Chichester (2010).
\bibitem{Ambjorn}J. Ambj{\o}rn, J. Jurkiewicz, and R. Loll, Phys. Rev. D {\bf 72}, 064014 (2005).
\bibitem{Podlubny}I. Podlubny, {\it Fractional Differential Equations\/}, Academic Press, New York (1998).
\bibitem{Thirring}W. E. Thirring, {\it A Course in Mathematical Physics\/}, vol. 3, Springer, Berlin (1981).
\bibitem{Amrein}W. O. Amrein, {\it Hilbert Space Methods in Quantum Mechanics\/}, EPFL Press, Lausanne (2009).
\bibitem{Edgar}G. Edgar, {\it Measure, Topology, and Fractal Geometry\/}, 2nd edition, Springer, New York (2008).
\bibitem{Nottale}L. Nottale, {\it Scale Relativity and Fractal Space-Time\/}, Imperial College Press, London (2011).
\bibitem{Rudin}W. Rudin, {\it Principles of Mathematical Analysis\/}, 3rd edition, McGraw-Hill, New York (1976).
\bibitem{Hilger}S. Hilger, Results Math. {\bf 18}, 19 (1990).
\bibitem{Bohner}M. Bohner and A. Peterson, {\it Dynamic Equations on Time Scales\/}, Birkh\"auser, Boston (2001).
\bibitem{Cieslinski}J. L. Cie\'sli\'nski, J. Math. Anal. Appl. {\bf 388}, 8 (2012).
\bibitem{Tsallis}C. Tsallis, {\it Introduction to Nonextensive Statistical Mechanics\/}, Springer, London (2009).
\bibitem{Naudts}J. Naudts, {\it Generalised Thermostatistics\/}, Springer, London (2011).
\bibitem{CN}M. Czachor and J. Naudts, Phys. Lett. A {\bf 298}, 369 (2002).
\bibitem{Jizba}P. Jizba and T. Arimitsu, Ann. Phys. {\bf 312},  17–59 (2004).
\bibitem{Renyi}A. R\'enyi, MTA III. Oszt. K\"ozl. {\bf 10}, 251 (1960). Reprinted in {\it Selected Papers of Alfred R\'enyi\/}, vol. 2, pp. 526--552, Akad\'emiai Kiad\'o, Budapest (1976).
\bibitem{CHSH}J. F. Clauser et al., Phys. Rev. Lett. {\bf 23}, 880 (1969); Erratum Phys. Rev. Lett. {\bf 24}, 549 (1970).
\bibitem{FractalQM}C. P. Dettmann and N. E. Frankel, J. Phys. A {\bf 26}, 1009 (1993).
\bibitem{ACK}D. Aerts, M. Czachor, and M. Kuna, Crystallization of space: Space-time fractals from fractal arithmetics, arXiv:1506.00487 [gr-qc] (2015).
\bibitem{Benedetti}D. Benedetti, Phys. Rev. Lett. {\bf 102}, 111303 (2009).
\bibitem{Modesto}L. Modesto and P. Nicolini, Phys. Rev. D {\bf 81}, 104040 (2010).
\bibitem{Nicolini}P. Nicolini and E. Spallucci, Phys. Lett. B {\bf 695}, 290 (2011).
\bibitem{Calcagni}G. Calcagni, D. Oriti, and J. Thürigen, Phys. Rev. D {\bf 91}, 084047 (2015).
\bibitem{DG1}H.-D.~Doebner and G.~A.~Goldin, Phys.~Rev.~A {\bf 54}, 3764 (1996).
\bibitem{DG2}H.-D.~Doebner and G.~A.~Goldin,  Phys.~Lett.~A, {\bf 162}, 397 (1992); J.~Phys.~A: Math. Gen. {\bf 27}, 1771 (1994).
\bibitem{DG3}G.~A.~Goldin, Nonlinear Math. Phys. {\bf 4}, 7 (1997).
\bibitem{DGMC}M. Czachor, Phys. Rev. A 57, R2263 (1998).

\end{thebibliography}
\end{document}